\documentclass[prd,showpacs,preprint,nofootinbib]{revtex4}
\usepackage{graphicx,color,amsmath}
\usepackage{slashed}

\begin{document}
\title{SU(3)-symmetry breaking effects and mass splitting
in scalar and pseudoscalar $D$ mesons from QCD sum rules}
\author{Hong-Ying Jin$^1$, Jin Zhang$^1$ , Zhu-Feng Zhang$^2$ and T.G. Steele$^3$}
\affiliation{$^1$Institute of Modern Physics, Zhejiang University,
Hangzhou, Zhejiang, China\\
$^2$Department of Physics, Ningbo University, Ningbo, Zhejiang,
China\\
$^3$Department of Physics and Engineering Physics, University of
Saskatchewan, Saskatoon, Saskatchewan, Canada S7N 5E2}
\date{June 2009}

\begin{abstract}
Motivated by the similar mass splitting in light-light and
heavy-light $J^{P}=0^{-}$ and $J^{P}=0^{+}$ mesons, the
SU(3)-symmetry breaking effects splitting the massses in the $0^{-}$
and $0^{+}$ channels of the $D$ meson are analyzed in the framework
of QCD sum rules with an underlying  $c\bar{q}$ structure. We take
into account  operator mixing to obtain an infrared stable OPE
including complete  non-perturbative and perturbative $\mathcal
{O}(m_{q})$ corrections to the correlation function.  With the same
threshold for both channels, the mass splitting arising from the
sum-rules has the same behavior as the observed spectrum. In
particular, we obtain $m_{D_{s}}-m_{D_{d}}\sim35\rm{MeV}$ in the
$0^{-}$ channel and $m_{D_{d}}-m_{D_{s}}\sim12\rm{MeV}$ in the
$0^{+}$ channel at a renormalization scale $\mu=1\rm{GeV}$. The
splitting can be attributed to the different roles of mass effects
and the parity-dependent `` force '' induced from non-perturbative
QCD vacuum. Further analysis shows that due to this
``parity-dependent'' force it is natural that the mass gap of the
two states in the $0^{-}$ channel is larger than the $0^{+}$
channel. When we increase the renormalization scale  to
$\mu=1.3\rm{GeV}$ the splitting remains unchanged which demonstrates
a correct scale invariance. Combined with HQET, generalization to
other channels of charmed mesons and $b$-systems are briefly
discussed .

\end{abstract}


\maketitle
\newpage
\section{introduction}
The SU(3) quark model of hadrons\cite{gell-mann, zweig} provide an
intuitive understanding of hadronic properties. Due to the
non-perturbative nature of low energy QCD we have to employ
non-perturbative methods in the hadronic sector. The QCD sum rule
approach\cite{shifman, reinders, novikov} has proven to be a
successful non-perturbative method to extract reasonable results in
the hadronic sector. Similar success have been achieved with
light-cone QCD sum rules\cite{balitsky, chernyak} which represent a
further-developed version of the original sum-rule approach.
 We will not dwell on the overall success of the QCD sum rules,
 but will focus
on the light and heavy $J^{P}=0^{-}$ and $J^{P}=0^{+}$ channels.

It is  observed\cite{amsler} that the mass splitting in the
$J^{P}=0^{-}$ channel for the lowest light mesons is in line with
their underlying structures from the  naive quark model estimates.
This splitting of lowest light pseudoscalars with quantum numbers of
the $\pi$, $K$, $\eta$ and $\eta'$ is well accommodated in QCD sum
rules if the instanton effects are appropriately\cite{shuryak}
included since the instanton contributions to the correlation
function are different from each member of a multiplet due to its
dependence on the isospin and effective mass
$m^{\ast}_{q}\footnote{One should not confuse it with the ``
effective mass\,'' in the consituent quark model.}$. However, the
splitting in light scalar meson $J^{P}=0^{+}$ is
 the reverse of  the naive
quark model estimate. If instanton effects are considered, the
splitting in the $J^{P}=0^{+}$ channel above 1GeV (i.e.
$f_{0}(1370)$, $a_{0}(1450)$, $K_{0}^{\ast}(1430)$ and
$f_{0}(1500)$)can also be explained within the framework of the QCD
sum rule approach\footnote{The results show there is large glueball
content in $f_{0}(1500).$\cite{zhang}}.

If we assume an ordinary light-heavy underlying structure of
open-charm systems a similar mass hierarchy as light pseudoscalars
can also be observed in the $J^{P}=0^{-}$ channel of $D$ mesons, the
$D_{d}(1869)$ and $D_{s}(1968)$. However, in the $J^{P}=0^{+}$
channel $D_{s}(2317)$ (which was first discovered by BARBAR
Collaboration\cite{bar} and later confirmed by CLEO\cite{cleo}) and
its isospin partner $D^{\ast}(2400)^{+}$ (observed by FOCUS
Collaboration\cite{link}) also show similar splitting as light
scalars of $J^{P}=0^{+}$ channel in contradiction to the naive quark
model estimate. Among these open-charm systems $D_{s}(2317)$
triggers much attention on its underlying structure. Mass results
from Lattice QCD for $D_{s}(2317)$ are larger than the experimental
value\cite{bali, dougall, lin} and the results in Ref.\,\cite{bali}
suggested that $D_{s}(2317)$ might receive a large $DK$ component.
The work of Ref.\,\cite{ybdai} including this contribution from $DK$
continuum in QCD sum rules based on a $\bar{c}s$ structure found
that this continuum contribution can significantly lower the mass
and decay constant of the $D_{s}(2317)$. A more complete work on
open-charm systems can be found in\,\cite{haya} where the
$J^{P}=0^{-}, 0^{+}, 1^{-}, 1^{+}$ channels were studied from a
viewpoint of $c\bar{q}$ system where $q=u, d, s$. On the contrary,
in considering the difficulty of ordinary heavy-light structure in
decoding the nature of $D_{s}(2317)$, a four-quark states picture
was proposed\cite{hycheng, barnes}.Ref.\,\cite{bracco} employed the
four-quark structure to investigate $D_{s}(2317)$ using QCD sum
rules which suggested that $D_{s}(2317)$ might be a four-quark
states, while the radiative decay of $D_{s}(2317)$ in light-cone sum
rules favors a $\bar{c}s$ structure\cite{colangelo}. All the work
both experimentally and theoretically shed some light on the
interpretation of $D_{s}(2317)$.

One might hope that the instanton improved QCD sum rules can also
realize the splitting in pseudoscalar and scalar $D$ mesons. But
this seems unlikely because there will be rapid suppression of
instanton contributions by the c-quark effective mass $m^{\ast}_{c}$
and damping exponential factor in the case of the charm quark, and
thus the QCD vacuum condensates of various operators are still the
dominant non-perturbative corrections. These condensates do not
preserve an ideal SU(3) flavor symmetry because of the
symmetry-breaking effect of different quark masses.

In the framework of QCD sum rules the SU(3)-symmetry breaking
effects present itself mainly from two sides. One is the different
values of $\langle\bar{u}u\rangle$, $\langle\bar{d}d\rangle$ and
$\langle\bar{s}s\rangle$, and associated mixed condensates.
Ref.\cite{snarison} studied the role of the chiral condensate
$\langle\bar{q}q\rangle$ in the mass splitting between the
scalar-pseudoscalar $D$(and $B$) mesons with so-called \emph{no-free
parameters} sum rules in the chiral limit, and recently a parallel
analysis was applied to the dependence of heavy-baryons mass
splitting on the ratio
$\kappa=\langle\bar{s}s\rangle/\langle\bar{d}d\rangle$\cite{narison3}.
Another symmetry-breaking effect is the perturbative mass correction
which is proportional to $m_{q_{1}}m_{q_{2}}$ where $q_{1}$ and
$q_{2}$ labels the quark content of the meson or the current
considered. These breaking effects are small for pure-light mesons
since the quark condensates, mixing condensates as well as the
perturbative mass corrections are always accompanied by the light
quark masses and therefore they will be greatly suppressed,
especially for $m_{u}$ and $m_{d}$. In fact we always use a massless
approximation for pure light systems. However, when heavy quarks are
involved, the mass effects will be considerable since the large mass
of heavy quark takes the place of one of the light ones mentioned
above. In other words the large mass of the heavy quark in
heavy-light systems results in more significant   mass effects
 than the pure-light system.
Thus it is more consistent to take into account the mass corrections
to an uniform order. The work\cite{haya} went further by including
the perturbative mass corrections, and the splitting in different
states was realized by choosing different thresholds. However, the
theoretical splitting(given by the central value) in $0^{+}$
channel, although in agreement with the potential model
results\cite{pierro}, still contradicts experiment. Therefore a
complete analysis including both
 operator mixing effects and all the dimension-6 operators is
necessary.

Motivated by the important role of heavy quark, in this work we will
 use QCD sum-rules to investigate the mass effects in splitting the
pseudoscalar and scalar $D$ multiplet. To be specific, in our work
complete perturbative and no-prerturbative $\mathcal {O}(m_{q})$
mass corrections  (where $m_{q}$ is the mass of light quark) are
taken into account in the sum rules. Unlike pure-light meson
SU(3)-breaking effects mainly introduced by VEVs of
renormalization-group invariant mass-dependent operators, in
heavy-light system there will be another source introduced by
operator mixing mentioned above. Another important point is that in
QCD sum rules the threshold also represents SU(3)-symmetry breaking,
so different members of the same SU(3) multiplet should have
different thresholds.In practice we have no knowledge about
SU(3)-symmetry breaking effects \emph{a priori}; it is only evident
from experiments. For instance we even do not know how to set the
thresholds  for the $0^{+}$ doublet before we know their exact
masses experimentally. Normally if we take the viewpoint of
constituent quark, the threshold of $D_{s}$ should be larger than
that of $D_{d}$. But if the mass of $D_{d}$ is larger than that of
$D_{s}$ in $0^{+}$ channel it would be not reasonable to set a
larger threshold for $D_{s}$. For this reason we would like to focus
our attention on the splitting trend rather than the exact spectrum
of the $D$ mesons. We believe the mass splitting of the doublet
$D_{s}$ and $D_{d}$ in QCD sum rules under the same
``\,reference\,'' threshold reflects the SU(3)-splitting tendency.
Thus for an exact sum rule a suitable modification of threshold due
to SU(3)-breaking should enlarge this tendency, otherwise we think
it is not \emph{natural}. To this end in our analysis we first
select one member of a multiplet as ``\,benchmark\,'' to fix the
suitable threshold and Borel window. Then we apply these parameters
to  another member with replacing the SU(3)-breaking dependent
quantities.

The article is structured as follows. In Section II we first review
the necessary results on the operator mixing and cancelation of mass
singularities for heavy-light current, then present the sum rules
for pseudoscalar and scalar currents of $D^{+}$ mesons. In Section
III the numerical results and discussion will be given. Conclusions
are presented in Section IV.

\section{the formulas}

\subsection{operator mixing and cancelation of mass singularities}

In order to demonstrate the operator mixing and cancelation of mass
singularities in heavy-light quark system, we consider the following
charmed scalar two-point function:
\begin{eqnarray}
\Pi(q^{2})&=&i\int d^{4}xe^{iqx}\langle0|T\{\bar{q}(x)c(x),
\bar{c}(0)q(0)\}| 0\rangle\nonumber\,\\
&=&\Pi^{\rm{pert}}(q^{2})+\Pi^{\rm{np}}(q^{2})\nonumber\\
&=&C_{I}(q^{2})I+\sum_{d\neq 0}C_{d}(q^{2})\langle
0|O_{d}|0\rangle,\nonumber
\end{eqnarray}
where $d$ is the dimension of the operator and $q=u, d, s$. For
simplicity the renormalization invariant factor
$(\ln(\mu/\Lambda))^{-4/b}$ has been suppressed where $\mu$ is the
normalization point and $b=(11N_{c}-2n_{f})/3$. Setting aside the
perturbative part until later, the contributions of VEVs of $d\leq
6$ may be written as\cite{general}:
\begin{eqnarray}
\Pi^{\rm{np}}(q^{2})&=&\bar{C}_{G^{2}}\langle\frac{\alpha_{s}}{\pi}
G^{2}\rangle+\bar{C}_{G^{3}}\langle\frac{\alpha_{s}}{\pi}
G^{3}\rangle+\bar{C}_{j^{2}}\langle
j^{2}\rangle\nonumber\,\\
&&+\bar{C}_{\bar{c}c}\bigg\{\langle\bar{c}c\rangle+\frac{1}{12m_{c}}\langle\frac{\alpha_{s}}{\pi}
G^{2}\rangle
+\frac{1}{360m_{c}^{3}}\Big(\langle\frac{\alpha_{s}}{\pi}G^{3}\rangle+12\langle j^{2}\rangle\Big)\bigg\}\nonumber\,\\
&&+\bar{C}_{\bar{c}Gc}\bigg\{\langle\bar{c}Gc\rangle-\frac{m_{c}}{2}\ln\frac{m_{c}^{2}}{\mu^{2}}\langle\frac{\alpha_{s}}{\pi}
G^{2}\rangle +\frac{1}{12m_{c}}\Big(\langle\frac{\alpha_{s}}{\pi}
G^{3}\rangle+2\langle
j^{2}\rangle\Big)\bigg\}\nonumber\,\\
&&+(c\rightarrow q)\nonumber\,\\
&&+\bar{C}_{\bar{q}jq}\bigg\{\langle
\bar{q}jq\rangle-\frac{1}{24}\ln\frac{m_{q}^{2}}{\mu^{2}}\langle
j^{2}\rangle\bigg\}\label{scalarnp}
\end{eqnarray}
The operators in Eq.(\ref{scalarnp}) are defined as follows:
\begin{equation}
\langle\frac{\alpha_{s}}{\pi}G^{2}\rangle=\Big\langle\frac{\alpha_{s}}{\pi}G^{a}_{\mu\nu}G^{a\mu\nu}\Big\rangle\,\nonumber
\end{equation}
\begin{equation}
\langle\frac{\alpha_{s}}{\pi}G^{3}\rangle=\Big\langle
g_{s}f^{abc}G^{a}_{\mu\nu}G^{b}_{\nu\rho}G^{c}_{\rho\mu}\Big\rangle\nonumber
\end{equation}
\begin{equation}
\langle j^{2}\rangle=\Big\langle
g_{s}^{2}(D_{\mu}G^{a\rho\mu})(D^{\nu}G^{a}_{\rho\nu})\Big\rangle=\bigg\langle
g_{s}^{4}\Big(\sum_{q}\bar{q}\gamma^{\rho}T^{a}q\Big)^{2}\bigg\rangle\nonumber
\end{equation}
\begin{equation}
\langle\bar{q}jq\rangle=\Big\langle
g_{s}\bar{q}\gamma_{\mu}(D_{\nu}G^{a\mu\nu})T^{a}q\Big\rangle=\bigg\langle
g_{s}^{2}\Big(\bar{q}\gamma_{\mu}T^{a}q\Big)\sum_{q}\bar{q}\gamma^{\mu}T^{a}q\bigg\rangle\nonumber
\end{equation}
therefore the contribution from $\langle j^{2}\rangle$ is to
$\mathcal {O}(\alpha_{s}^{2})$, in the following we assume the
singularities in $j^{2}$ are canceled by the mixing and we will omit
this term. All the charmed-condensates vanish by virtue of
\emph{heavy quark expansion}\cite{shifman, general, ebagan, bagan}:
\begin{eqnarray}
\langle
\bar{c}c\rangle=-\frac{1}{12m_{c}}\langle\frac{\alpha_{s}}{\pi}G^{2}\rangle
-\frac{1}{360m_{c}}\langle\frac{\alpha_{s}}{\pi}G^{3}\rangle+...\nonumber\,\\
\langle
\bar{c}Gc\rangle=\frac{m_{c}}{2}\ln\frac{m_{c}^{2}}{\mu^{2}}\langle\frac{\alpha_{s}}{\pi}G^{2}\rangle
-\frac{1}{12m_{c}}\langle\frac{\alpha_{s}}{\pi}G^{3}\rangle+...\label{hqexpan}
\end{eqnarray}
thus there are only gluonic and light quark related condensates
left. It is clear there are mixing to gluonic operators from
$\langle\bar{q}q\rangle$ and $\langle\bar{q}Gq\rangle$ terms, and
with the help of this mixing the singular parts in gluonic
coefficients in limit $m_{q}\rightarrow0$ will be well canceled and
we are left
 with an infrared stable expression. The final-form of various non-perturbative
coefficients follows  Eq.(\ref{scalarnp}):
\begin{eqnarray}
C_{\bar{q}q}&=&\bar{C}_{\bar{q}q},\nonumber\,\\
C_{G^{2}}&=&\bar{C}_{G^{2}}+\frac{1}{12m_{q}}\bar{C}_{\bar{q}q}
-\frac{m_{q}}{2}\ln\frac{m_{q}^{2}}{\mu^{2}}\bar{C}_{\bar{q}Gq},\nonumber\,\\
C_{\bar{q}Gq}&=&\bar{C}_{\bar{q}Gq},\nonumber\,\\
C_{\bar{q}jq}&=&\bar{C}_{\bar{q}jq},\nonumber\,\\
C_{G^{3}}&=&\bar{C}_{G^{3}}
+\frac{1}{360m_{q}^{3}}\bar{C}_{\bar{q}q}+\frac{1}{12m_{q}}\bar{C}_{\bar{q}Gq}.\label{finalcoef}
\end{eqnarray}
Three of the $\bar{C}s$ have been worked out\cite{mjamin} in
expansion in $m_{q}$. In our notation:
\begin{eqnarray}
\bar{C}_{\bar{q}q}&=&-\frac{m_{c}}{q^{2}-m_{c}^{2}}
+\frac{m_{q}}{2}\frac{2m_{c}^{2}-q^{2}}{(q^{2}-m_{c}^{2})^{2}}-\frac{m_{q}^{2}m_{c}^{3}}{(q^{2}-m_{c}^{2})^{3}},\nonumber\,\\
\bar{C}_{G^{2}}&=&\frac{1}{12(q^{2}-m_{c}^{2})}\bigg\{\frac{m_{c}}{m_{q}}-\frac{q^{2}}{2(q^{2}-m_{c}^{2})}
+\frac{m_{q}q^{2}}{m_{c}(q^{2}-m_{c}^{2})^{2}}\Big[q^{2}+6m_{c}^{2}\nonumber\\
&&+6m_{c}^{2}\Big(\frac{1}{2}\ln\frac{m_{q}^{2}}{\mu^{2}}+\ln\frac{\mu
m_{c}}{m_{c}^{2}-q^{2}}\Big)\Big]
-\frac{3m_{q}^{2}m_{c}^{2}q^{2}}{(q^{2}-m_{c}^{2})^{3}}\ln\frac{m_{q}^{2}}{\mu^{2}}\bigg\},\nonumber\,\\
\bar{C}_{\bar{q}Gq}&=&\frac{m_{c}}{2}\frac{q^{2}}{(q^{2}-m_{c}^{2})^{3}}
-\frac{m_{q}m_{c}^{2}}{2}\frac{q^{2}}{(q^{2}-m_{c}^{2})^{4}},\label{unmixcoef}
\end{eqnarray}
and the singular pieces of $\bar{C}_{G^{3}}$ as $m_{q}\rightarrow 0$
are\footnote{The three-gluonic coefficient in \cite{ebagan}was
derived from pseudoscalar heavy-light current, here one can obtain
the scalar one by replacing $m_c$ by $-m_c$. }\cite{ebagan}:
\begin{equation}
\bar{C}_{G^{3}}=\frac{m_{c}}{360m_{q}^{3}(q^{2}-m_{c}^{2})}
+\frac{q^{2}-2m_{c}^{2}}{720m_{q}^{2}(q^{2}-m_{c}^{2})^{2}}
-\frac{m_{c}(15q^{2}-m_{c}^{2})}{360m_{q}(q^{2}-m_{c}^{2})^{3}}.\label{GGG}
\end{equation}
Substituting Eq.(\ref{unmixcoef}) and Eq.(\ref{GGG}) into
Eq.(\ref{finalcoef}) all the mass singular parts of $C_{G^{2}}$ and
$C_{G^{3}}$ appearing as $1/m_{q}$ and $\ln m_{q}$ are canceled
since these terms are remnants of long distance structure of vacuum
condensates. For definiteness we write down the explicit form of
$C_{G^{2}}$ to $\mathcal {O}(m_{q})$ as:
\begin{eqnarray}
C_{G^{2}}&=&-\frac{1}{12(q^{2}-m_{c}^{2})}
-\frac{m_{q}m_{c}^{3}}{12(q^{2}-m_{c}^{2})^{3}}\nonumber\,\\
&&+\frac{m_{q}q^{2}}{12m_{c}(q^{2}-m_{c}^{2})^{3}}\Big(q^{2}+6m_{c}^{2}+6m_{c}^{2}\ln\frac{\mu
m_{c}}{m_{c}^{2}-q^{2}}\Big),\label{finalGG}
\end{eqnarray}
Therefore to $\mathcal {O}(m_{q})$ the operator mixing changes the
$C_{G^{2}}$ significantly and it is expected there will be new mass
effects on the sum rules. Similarly, for $C_{G^{3}}$ we have(see the
appendix for details):
\begin{equation}
C_{G^{3}}=-\frac{q^{2}}{720m_{c}^{6}}W(-10W^{3}+4W^{2}+3W+2),\label{finalGGG}
\end{equation}
where
\begin{equation}
W=\frac{m_{c}^{2}}{m_{c}^{2}-q^{2}}.\nonumber
\end{equation}
Now we have fixed the non-perturbative parts in the OPE of scalar
heavy-light current. It is easily to get the non-perturbative parts
of pseudoscalar current by replacing $m_{c}$ by $-m_{c}$ in
Eq.(\ref{finalcoef}). The perturbative part and $C_{\bar{q}jq}$ will
be presented in the forthcoming subsection.

\subsection{the sum rules}

The sum rules of scalar and pseudoscalar $D$ are based on the
following two-point correlation function:
\begin{equation}
\Pi_{\Gamma}(q^2)=i\int d^4 x\,e^{iq
x}\langle0|\,T\{\bar{q}(x)\Gamma c(x), \, \bar{c}(0)\Gamma
q(0)\}|0\rangle,\label{corr}
\end{equation}
where $q$ is the light flavor in the $D$ meson,
$\Gamma=\{I,i\gamma_{5}\}$ for scalar and pseudoscalar $D$ meson
respectively. The decay matrix element of scalar $D$ meson is
defined as:
\begin{equation}
\langle 0|\bar{q}c|0\rangle=m_{D}f_{D},\nonumber
\end{equation}
and following\cite{khod} the pseudoscalar one is defined as:
\begin{equation}
m_{c}\langle0|\bar{q}i\gamma_{5}c|D\rangle=m_{D}^{2}f'_{D},\nonumber
\end{equation}
where $m_{c}$ is the c-quark mass and $m_{D}$ the $D$ mass, and
$m_{q}$ labels the mass of light quark. Including the perturbative
mass correction to $\mathcal {O}(m_{q})$ and to $O(\alpha_{s})$ as
well as the complete $\mathcal {O}(m_{q})$ non-perturbative
corrections worked out in previous subsection, after Borel
transformation the OPE of correlation function in Eq.(\ref{corr}) is
given by\footnote{The $m_{q}$-independent term in coefficient of
$\langle\frac{\alpha_{s}}{\pi}G^{2}\rangle$ here is different
from\cite{haya} and\cite{snarison}, but this term has no impact on
splitting except an uniform shift.}\cite{ljreinders, aliev, khod}:
\begin{eqnarray}
\Pi_{\Gamma}^{{\rm{OPE}}}(M^{2})&=&\frac{3}{8\pi^{2}M^{2}}\int_{m_{c}^{2}}^{\infty}ds\frac{(s-m_{c}^{2})^{2}}{s}
\bigg[1\mp\frac{2m_{q}m_{c}}{s-m_{c}^{2}}
+\frac{4\alpha_{s}}{3\pi}f(s,m_{c}^{2})\bigg]\exp[-\frac{s}{M^{2}}]\nonumber\\
&&+\bigg[\pm\frac{m_{c}}{M^{2}}+\frac{1}{2M^{2}}\Big(1+\frac{m_{c}^{2}}{M^{2}}\Big)m_{q}\bigg]\langle\bar{q}q\rangle\exp[-\frac{m_{c}^{2}}{M^{2}}]\,\nonumber\\
&&+\Bigg\{\frac{1}{12M^{2}}\pm\bigg[\frac{m_{q}m_{c}^{3}}{8M^{6}}
+\frac{m_{q}m_{c}}{2M^{4}}\Big(1-\frac{m_{c}^{2}}{2M^{2}}\Big)\Big(\gamma_{E}+\ln\frac{\mu m_{c}}{M^{2}}\Big)\,\nonumber\\
&&-\frac{m_{q}}{12m_{c}M^{2}}\Big(1-\frac{2m_{c}^{2}}{M^{2}}\Big)\bigg]\Bigg\}\langle\frac{\alpha_{s}}{\pi}G^{a}_{\mu\nu}G^{a\mu\nu}\rangle\exp[-\frac{m_{c}^{2}}{M^{2}}]\,\nonumber\\
&&+\Bigg\{\pm\frac{1}{2M^{4}}\Big(1-\frac{m_{c}^{2}}{2M^{2}}\Big)
m_{c}\langle\bar{q}g\sigma_{\mu\nu}\frac{\lambda^{a}}{2}G^{a\mu\nu}q\rangle\,\nonumber\\
&&+\frac{m_{c}^{2}}{4M^{6}}\Big(1-\frac{m_{c}^{2}}{3M^{2}}\Big)m_{q}\langle\bar{q}g\sigma_{\mu\nu}\frac{\lambda^{a}}{2}G^{a\mu\nu}\rangle\,\nonumber\\
&&-\frac{16\pi\alpha_{s}}{27M^{4}}\Big(1-\frac{m_{c}^{2}}{4M^{2}}-\frac{m_{c}^{4}}{12M^{4}}\Big)
\langle\bar{q}q\rangle^{2}\Bigg\}\exp[-\frac{m_{c}^{2}}{M^{2}}],\label{OPE}
\end{eqnarray}
where
\begin{eqnarray}
f(s,m_{c}^{2})&=&\frac{9}{4}+2{\rm{Li}}_{2}(\frac{m_{c}^{2}}{s})
+\ln\frac{s}{m_{c}^{2}}\ln\frac{s}{s-m_{c}^{2}}
+\frac{3}{2}\ln\frac{m_{c}^{2}}{s-m_{c}^{2}}\nonumber\\
&&+\ln\frac{s}{s-m_{c}^{2}}+\frac{m_{c}^{2}}{s}\ln\frac{s-m_{c}^{2}}{m_{c}^{2}}
+\frac{m_{c}^{2}}{s-m_{c}^{2}}\ln\frac{s}{m_{c}^{2}},\nonumber
\end{eqnarray}
with
\begin{equation}
{\rm{Li_{2}}}(x)=-\int_{0}^{x}dt\,\frac{\ln(1-t)}{t},\nonumber
\end{equation}
and
\begin{equation}
\gamma_{E}=0.577.\nonumber
\end{equation}
is the Euler constant. Contributions from 3-gluonic condensates have
been omitted safely since it is $m_{q}$-independent and greatly
suppressed by the huge denominator therefore it is not responsible
for the splitting as one can see from Eq.(\ref{finalGGG}). The upper
and lower signs in Eq.(\ref{OPE}) are for the scalar and
pseudoscalar channel respectively.

On the other hand the correlation function in Eq.(\ref{corr}) can
also be derived from the phenomenological side by the dispersion
relation:
\begin{equation}
\Pi_{\Gamma}(q^{2})=\frac{1}{\pi}\int_{m_{c}^{2}}^{\infty}ds\,\frac{{\rm{Im}}\Pi_{\Gamma}^{{\rm{ph}}}(s)}{s-q^{2}}
+\rm{subtraction\,\,constant},\label{disp}
\end{equation}
where the spectral density $\rm{Im}\Pi_{\Gamma}^{ph}(s)$ is obtained
by inserting a complete set of quantum states $\Sigma |n\rangle
\langle n|$ into Eq.(\ref{corr}) which reads:
\begin{equation}
{\rm{Im}}\Pi_{\Gamma}(s)=F_{\Gamma}m_{D}^{2}\pi\,\delta(s-m_{D}^{2})+\pi\frac{3}{8\pi^{2}}
\frac{(s-m_{c}^{2})^{2}}{s}\Big[1\mp\frac{2m_{q}m_{c}}{s-m_{c}^{2}}
+\frac{4\alpha_{s}}{3\pi}f(s,m_{c}^{2})\Big]\theta(s-s_{0}),\label{ph}
\end{equation}
Taking the Borel transformation of Eq.(\ref{disp}) and equating it
with Eq.(\ref{OPE}), after subtracting the continuum contributions
we arrive the desired sum rules:
\begin{eqnarray}
F_{\Gamma}m_{D}^{2}\exp[-\frac{m_{D}^{2}}{M^{2}}]&=&\frac{3}{8\pi^{2}}\int_{m_{c}^{2}}^{s_{0}}ds\,
\frac{(s-m_{c}^{2})^{2}}{s}\bigg[1\mp\frac{2m_{q}m_{c}}{s-m_{c}^{2}}
+\frac{4\alpha_{s}}{3\pi}f(s,m_{c}^{2})\bigg]\exp[-\frac{s}{M^{2}}]\,\nonumber\\
&&+\bigg[\pm m_{c}\langle\bar{q}q\rangle+\frac{1}{2}\Big(1+\frac{m_{c}^{2}}{M^{2}}\Big)m_{q}\langle\bar{q}q\rangle\bigg]\exp[-\frac{m_{c}^{2}}{M^{2}}]\,\nonumber\\
&&+\Bigg\{\frac{1}{12}\pm\bigg[\frac{m_{q}m_{c}^{3}}{8M^{4}}
+\frac{m_{q}m_{c}}{2M^{2}}\Big(1-\frac{m_{c}^{2}}{2M^{2}}\Big)\Big(\gamma_{E}+\ln\frac{\mu m_{c}}{M^{2}}\Big)\,\nonumber\\
&&-\frac{m_{q}}{12m_{c}}\Big(1-\frac{2m_{c}^{2}}{M^{2}}\Big)\bigg]\Bigg\}\langle\frac{\alpha_{s}}{\pi}G_{\mu\nu}^{a}G^{a\mu\nu}\rangle\exp[-\frac{m_{c}^{2}}{M^{2}}]\,\nonumber\\
&&+\Bigg\{\pm\frac{1}{2M^{2}}\Big(1-\frac{m_{c}^{2}}{2M^{2}}\Big)m_{c}\langle\bar{q}g\sigma_{\mu\nu}\frac{\lambda^{a}}{2}G^{a\mu\nu}q\rangle\,\nonumber\\
&&+\frac{m_{c}^{2}}{4M^{4}}\Big(1-\frac{m_{c}^{2}}{3M^{2}}\Big)m_{q}\langle\bar{q}g\sigma_{\mu\nu}\frac{\lambda^{a}}{2}G^{a\mu\nu}\rangle\,\nonumber\\
&&-\frac{16\pi\alpha_{s}}{27M^{2}}\Big(1-\frac{m_{c}^{2}}{4M^{2}}
-\frac{m_{c}^{4}}{12M^{4}}\Big)\langle\bar{q}q\rangle^{2}\Bigg\}\exp[-\frac{m_{c}^{2}}{M^{2}}].\label{smr}
\end{eqnarray}

Now we have completed the sum rules for pseudoscalar and scalar $D$
mesons. The input parameters in Eq.(\ref{smr}) are as
follows\,\cite{khodj, cheng, narison, amsler}:
\begin{eqnarray}
\Lambda_{\rm{QCD}}=259\rm{MeV},\quad\alpha_{s}=0.517, \quad
&\langle\alpha_{s}G^{2}\rangle=(0.07\pm0.01)\rm{GeV^{4}},
\nonumber\\
\langle\bar{u}u\rangle=\langle\bar{d}d\rangle=-(0.225\pm0.025)^{3}\rm{GeV^{3}},
&\langle\bar{s}s\rangle=0.8\langle\bar{u}u\rangle,
\nonumber\\
m_{d}=3\sim7\rm{MeV},
&m_{s}=120{\rm{MeV}},\nonumber\\
\langle\bar{q}\sigma_{\mu\nu}\frac{\lambda^{a}}{2}G^{a\mu\nu}q\rangle=m_{0}^{2}\langle\bar{q}q\rangle,&
m_{0}^{2}=0.8{\rm{GeV^{2}}}.\label{param}
\end{eqnarray}
All the values adopted above are given at the scale $\mu=1\rm{GeV}$
and we deduce the QCD scale $\Lambda_{\rm{QCD}}$ to one-loop from
$\alpha_{s}(M_{Z})=0.1170\pm0.0012$\cite{mason}. The renormalization
scale dependence is given by\cite{cheng}:
\begin{eqnarray}
m_{q}(\mu)&=&m_{q}(\mu_{0})\Big(\frac{\alpha_{s}(\mu_{0})}{\alpha_{s}(\mu)}\Big)^{-4/b},\nonumber\\
\langle\bar{q}q\rangle(\mu)&=&\langle\bar{q}q\rangle(\mu_{0})\Big(\frac{\alpha_{s}(\mu_{0})}{\alpha_{s}(\mu)}\Big)^{4/b},\nonumber\\
\langle g_{s}\bar{q}\sigma Gq\rangle(\mu)&=&\langle
g_{s}\bar{q}\sigma
Gq\rangle(\mu_{0})\Big(\frac{\alpha_{s}(\mu_{0})}{\alpha_{s}(\mu)}\Big)^{-2/3b},\nonumber\\
\langle\frac{\alpha_{s}}{\pi}G^{2}\rangle(\mu)&=&\langle\frac{\alpha_{s}}{\pi}G^{2}\rangle(\mu_{0}).\label{rge}
\end{eqnarray}
with $b=(11N_{c}-2n_{f})/3$. We use the following pole mass for the
charm quark\footnote{A note in\cite{snarison} argued that the value
$m_{c}=1.46\rm{GeV}$ used in\cite{haya} may be ill-defined. As the
sum rule is sensitive to $m_{c}$ as we can see in the next section,
the larger choice $m_{c}=1.47\pm0.04\rm{GeV}$ here might induce new
error. But it does not affect the splitting since the thresholds are
fixed for same channel in our analysis.}\cite{narison2}:
\begin{equation}
m_{c}=1.47\pm0.04{\rm{GeV}},\nonumber\\
\end{equation}
which can be expressed in terms of the running mass through the
relation:
\begin{equation}
m_{c}=\bar{m}_{c}(\mu)\Big[1+\Big(\frac{4}{3}+\ln\frac{\mu^{2}}{m_{c}^{2}}\Big)\frac{\bar{\alpha}_{s}(\mu)}{\pi}
+\mathcal{O}(\alpha_{s}^{2})\Big].\label{polemass}
\end{equation}

Taking the logarithm of both sides of Eq.(\ref{smr}) and applying
the differential operator $M^4\partial/\partial M^2$ to them we can
separate the mass from decay constant. Now we have fixed all the
ingredients for numerical analysis.

\section{results and discussion}

Firstly we present the criteria followed in our analysis:

1. To specify the appropriate threshold and Borel window, we demand
that the continuum contribution [i.e., the part in the dispersive
integral from $s_{0}$ to $\infty$ which has been subtracted from
both sides of Eq.(\ref{smr})] should not be too large (less than
30\% of the total dispersive integral). This criterion give us an
upper limit on the Borel momentum $M^{2}$. Furthermore, the
non-perturbative dimension-six operators corrections should be less
than 10\% which establishes a lower limit.

2. In order to check the mass effects on the splitting of the same
$J^{P}$ channel, states of that channel will be analyzed under same
threshold while the flavor-dependent parameters such as $m_{q}$,
$\langle\bar{q}q\rangle$ as well as $\langle\bar{q}\sigma Gq\rangle$
change correspondingly. This will supply us with an appropriate
comparison in the same channel with different light content.

3. As mentioned above  our primary concern is a correct splitting
trend, not the whole spectrum in same channel. Thus we select one
state as our ``benchmark'' to determine the threshold and Borel
window according to criterion 1. For definiteness we select $D_{d}$
and $D_{s}$ as our sample in $0^{-}$ and in $0^{+}$ channel
respectively. After this we turn on another state following
criterion 2. If it is a \emph{natural} sum rule, it should produce a
correct splitting trend that agrees with the experiment.

With these criteria in mind we plot the mass curves of the two
states in the same channel against the Borel momentum $M^{2}$ in a
diagram for different threshold and charm mass $m_{c}$ since it is
convenient to observe the splitting. The working windows which
satisfy the criterion 1 are marked by two short lines(or one short
line which labels the upper limit only) while the narrow ranges from
which we read our numerical value  are marked by shaded bars. If
there is no an obvious extremum within the window we determined, the
central value will be adopted. Under these criteria we find for
fixed threshold and charm mass $m_{c}$ as well as scale parameter
$\mu$ the working windows for $D_{d}$ and $D_{s}$ in each channel
are very close. The upper limit of our working windows decrease as
the thresholds decrease, while it seems that the lower limit is
nearly invariant which can be seen from the following graphs. When
we scale up to $\mu=1.3\rm{GeV}$ the upper limit increases compared
with $\mu=1\rm{GeV}$ while there is no obvious impact on the lower
limit.

\begin{figure}
\begin{center}
\includegraphics[scale=0.75]{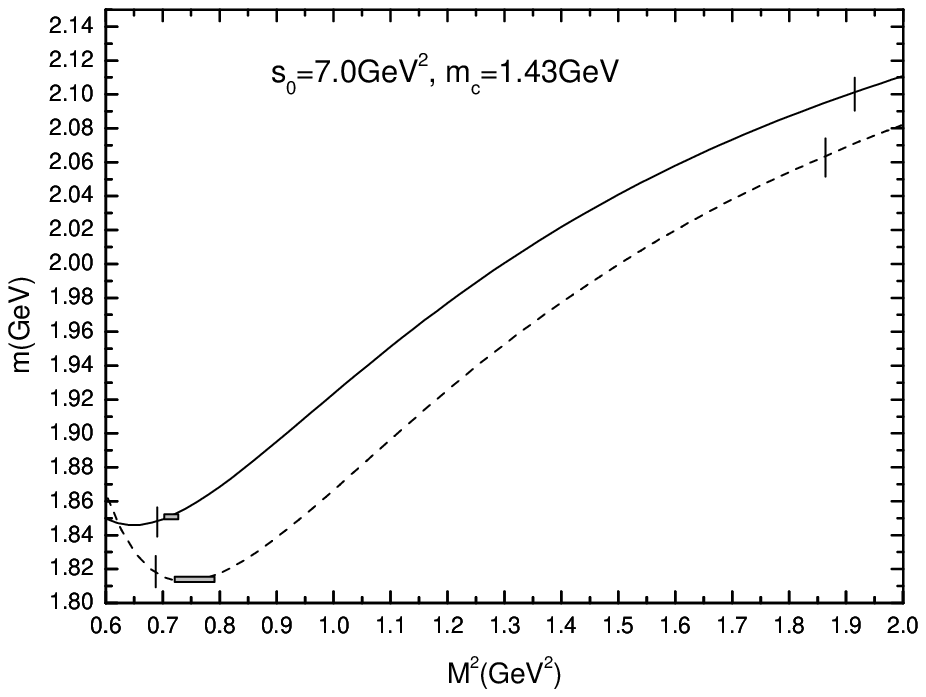}
\includegraphics[scale=0.75]{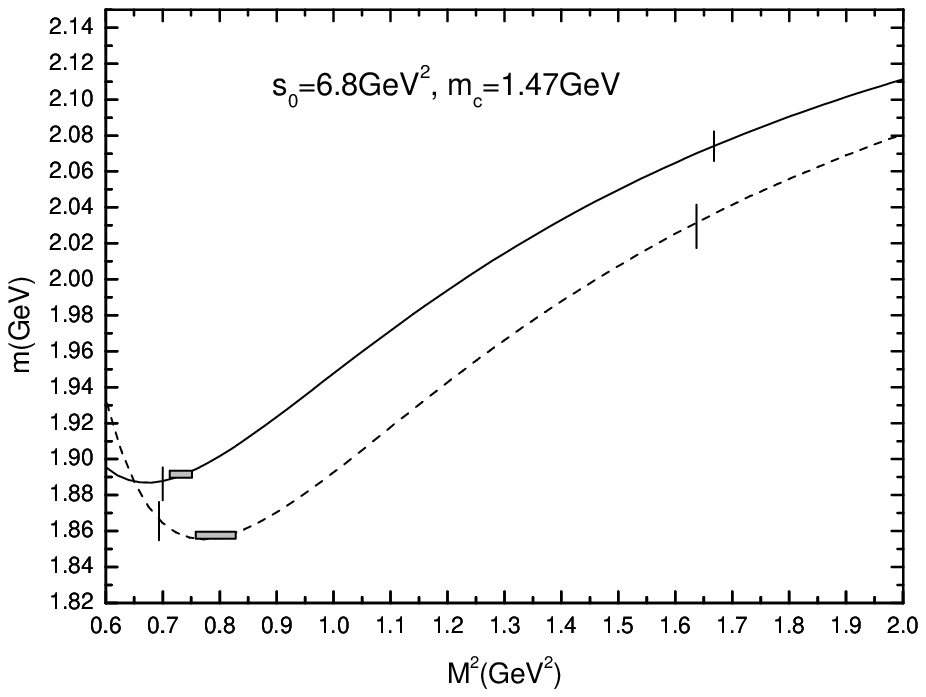}
\includegraphics[scale=0.75]{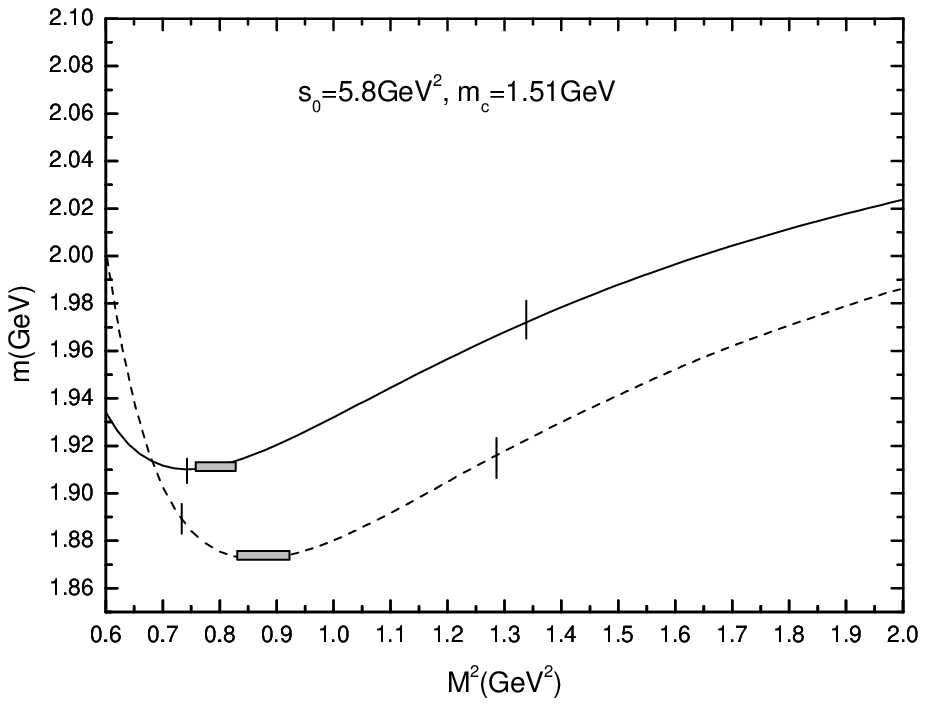}
\caption{Mass of $D_{d}$(dashed line) and $D_{s}$ meson (solid line)
of $0^{-}$ channel from pseudoscalar sum rule of Eq.(\ref{smr}) as
function of Borel momentum $M^{2}$ at scale $\mu=1\rm{GeV}$.}\label{ps}
\end{center}
\end{figure}

\begin{figure}
\begin{center}
\includegraphics[scale=0.75]{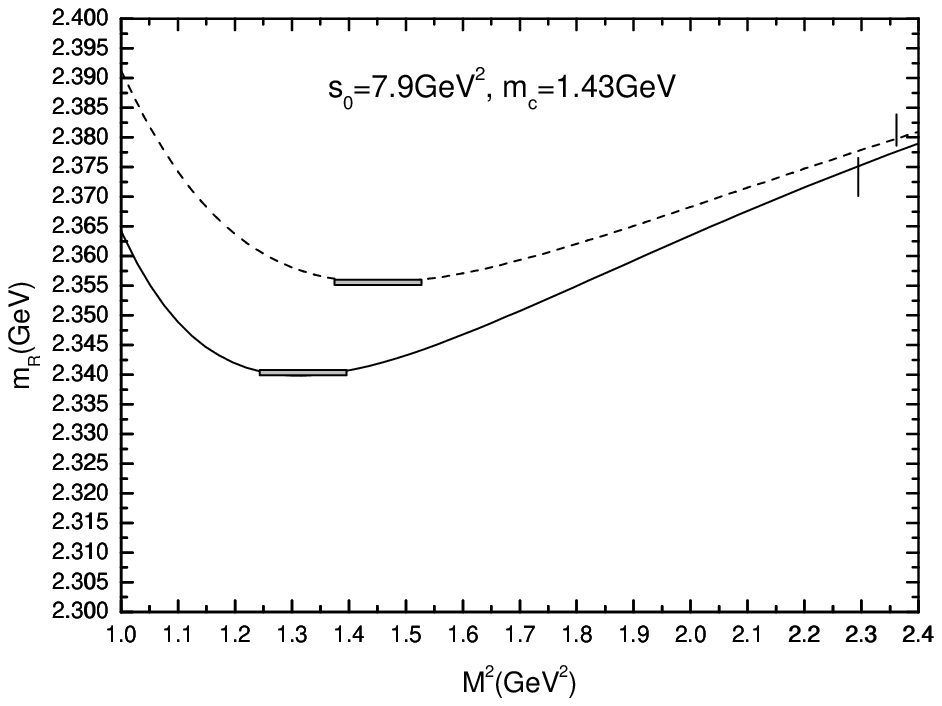}
\includegraphics[scale=0.75]{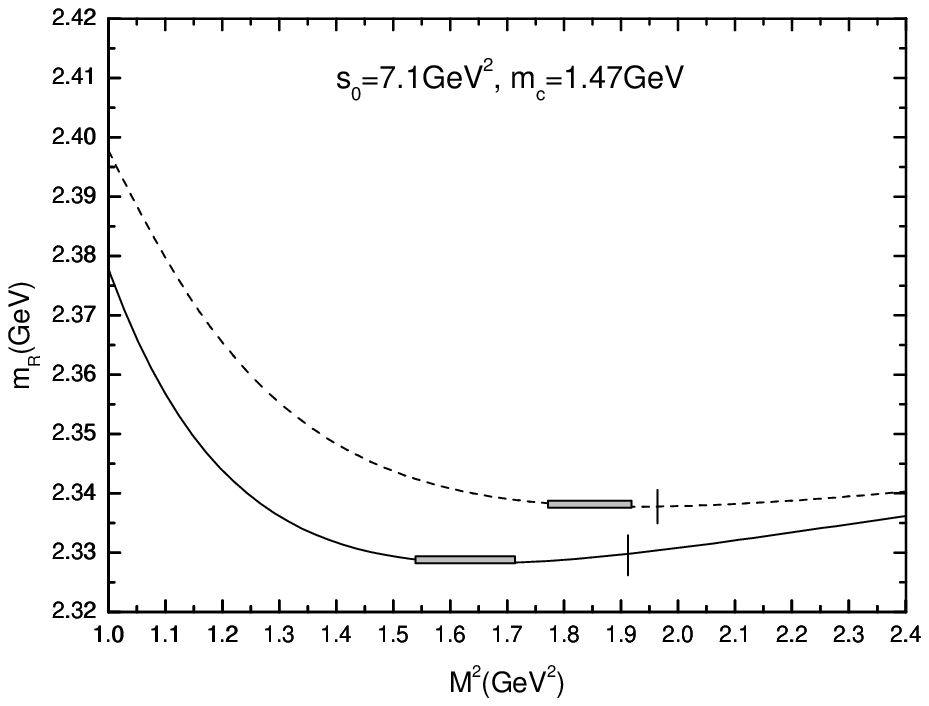}
\includegraphics[scale=0.75]{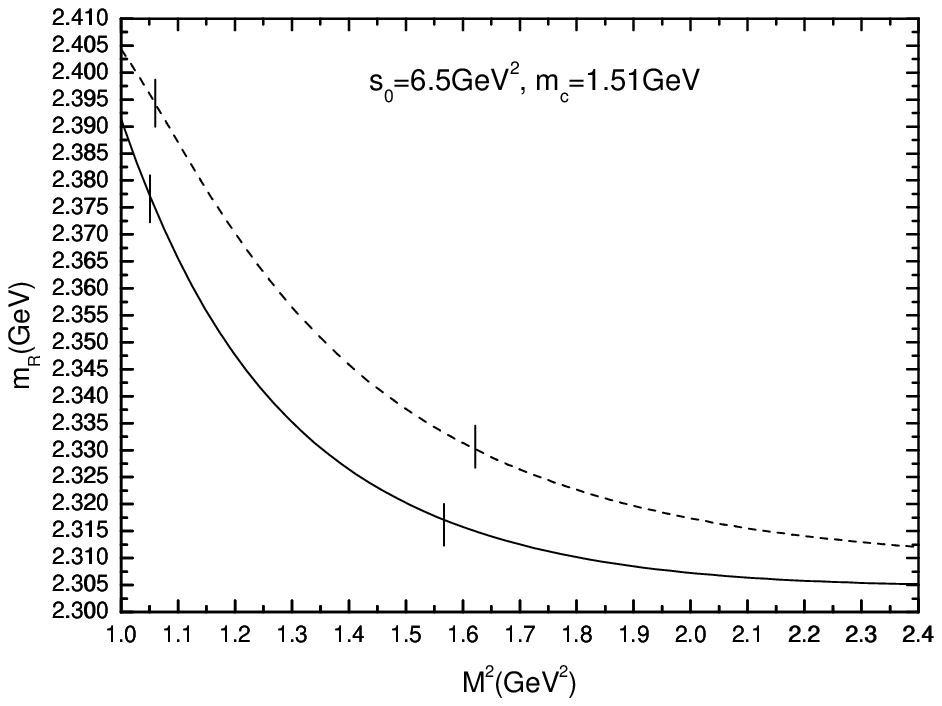}
\caption{Mass of $D_{d}$(dashed line) and $D_{s}$ meson (solid line)
from scalar sum rule of Eq.(\ref{smr}) as function of Borel momentum
$M^{2}$ at $\mu=1\rm{GeV}$.}\label{sc}
\end{center}
\end{figure}

First, we study the  $0^{-}$ channel from the pseudoscalar sum rule
given in Eq.(\ref{smr}) at the scale $\mu=1\rm{GeV}$. The numerical
results of the two states of $J^{P}=0^{-}$ channel are shown in
figure.\ref{ps}. We can see from figure.\ref{ps} that the results
following the pseudoscalar sum rules of Eq.(\ref{smr}) accurately
reproduce the mass of $D_{d}$ with a large charm mass which is very
close to the experimental $D_{d}(1869)$. When the parameters of
$D_{s}$ turn on and with the threshold fixed, the resulting mass
from the sum rule Eq.(\ref{smr}) is still lower compared with the
experimental one $D_{s0}(1968)$ but it is always larger than
$D_{d}$. The mass splitting from  Eq.(\ref{smr}) is
$\sim35\rm{MeV}$, a value much lower than the observed splitting
$\sim100\rm{MeV}$. At a first glance it appears that the
SU(3)-breaking effects can not supply a realistic splitting. But it
is important that the sum rule does present a reasonable mass
splitting trend in the $0^{-}$ channel which is in line with the
observed spectrum. The failure of pseudoscalar sum rules of
Eq.(\ref{smr}) in providing the entire mass gap in $0^{-}$ channel
is understandable: there is a large gap between the two $0^{-}$
states thus a threshold appropriate to $D_{d}(1869)$ is too low to
produce $D_{s}(1986)$. On the other hand in QCD sum rules we notice
the difference of threshold between different members also reflects
SU(3)-symmetry breaking. Therefore it is expected that if we
determine the threshold of each member separately, at some lager
threshold than the one for $D_{d}$ the mass of $D_{s}$ will be well
produced from the sum rules. In addition,
 the theoretical results are very sensitive to charm
mass: as we can see from figure.\ref{ps}, at the lowest value
adopted in our work, even at a much higher $s_{0}$ it is still
difficult to produce $D_{d}$. We can read from figure.\ref{ps} that
it seems the pole mass $m_{c}=1.47\rm{GeV}$ is more appropriate than
the other two choices. The results are summarized in
Table.\ref{tabps}.

\begin{table}[t]
\caption{Mass of pseudoscalar $D_{d}$ and $D_{s}$ read from shaded
areas marked by short bars in figure.\,\ref{ps} for different
threshold $s_{0}$ and $m_{c}$ at scale $\mu=1\rm{GeV}$. }
\begin{ruledtabular}
\begin{tabular}{c c c c c}
$s_{0}({\rm{GeV^{2}}})$ & $m_{c}(\rm{GeV})$ & $D_{d}(\rm{MeV})$ & $D_{s}(\rm{MeV})$ & $M_{D_{s}}-M_{D_{d}}(\rm{MeV})$ \\
\hline
 $7.0$ & $1.43$ & $1812$ & $1846$ & $34$\\
$6.8$ & $1.47$ & $1856$ & $1888$  & $32$\\
$5.8$ & $1.51$ & $1873$ & $1911$  & $38$\\
\end{tabular}
\end{ruledtabular}\label{tabps}
\end{table}

\begin{table}[t]
\caption{Masses of scalar $D_{d}$ and $D_{s}$ read from
figure.\,\ref{sc} for different threshold $s_{0}$ and $m_{c}$ at
scale $\mu=1\rm{GeV}$. The first two values are read from the shaded
area marked by short bars while the third is the central value
between the two short lines in the third graph.}
\begin{ruledtabular}
\begin{tabular}{c c c c c}
$s_{0}({\rm{GeV^{2}}})$ & $m_{c}(\rm{GeV})$ & $D_{d}(\rm{MeV})$  & $D_{s}(\rm{MeV})$ & $M_{D_{d}}-M_{D_{s}}(\rm{MeV})$\\
\hline
 $7.9$ & $1.43$ & $2356$ & $2340$ & $16$\\
$7.1$ & $1.47$ & $2341$ & $2328$ & $13$ \\
$6.5$ &$1.51$ & $2354$ & $2340$ & $14$ \\
\end{tabular}
\end{ruledtabular}\label{tabsc}
\end{table}

Now we turn to the analysis of $0^{+}$ channel from scalar sum rule
of Eq.(\ref{smr}). The observed spectrum of $0^{+}$ channel is
reversed from   the estimate from naive quark model. The numerical
results following scalar sum rules of Eq.(\ref{smr}) are shown in
figure.\ref{sc}. We can see that under the threshold determined from
$D_{s}$, a mass gap $M_{D_{d}}-M_{D_{s}}\sim15\rm{MeV}$ between
$D_{d}$ and $D_{s}$ can be realized which is also lower than the
experimental one $\sim35\rm{MeV}$. However, it indeed gives a
correct splitting trend which agrees with experiment. The results in
the $0^{+}$ channel are summarized in Table.\ref{tabsc}.

\begin{figure}
\begin{center}
\includegraphics[scale=0.75]{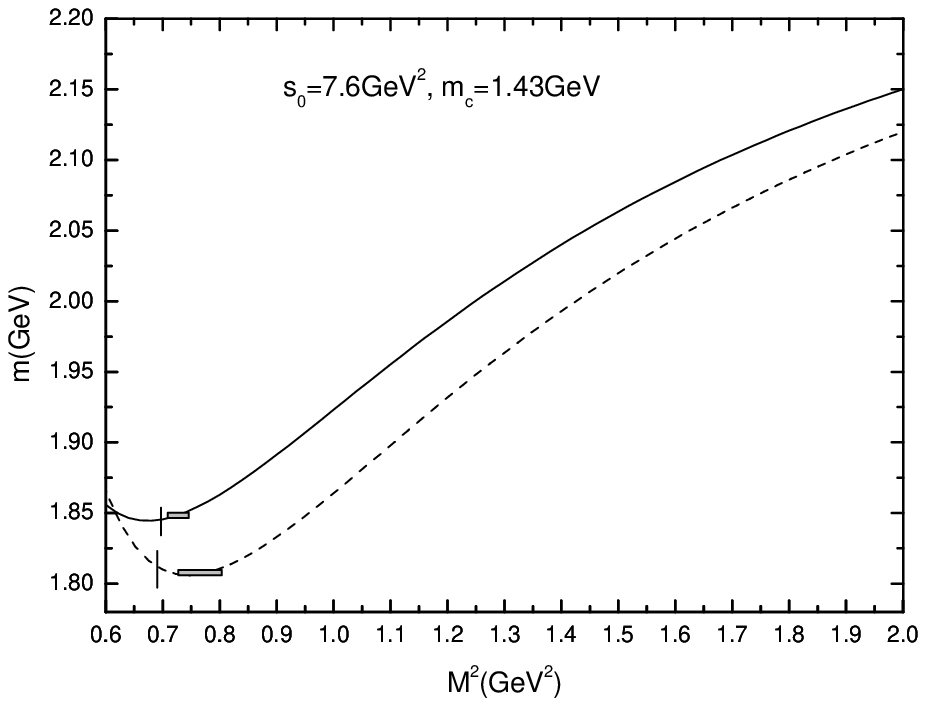}
\includegraphics[scale=0.75]{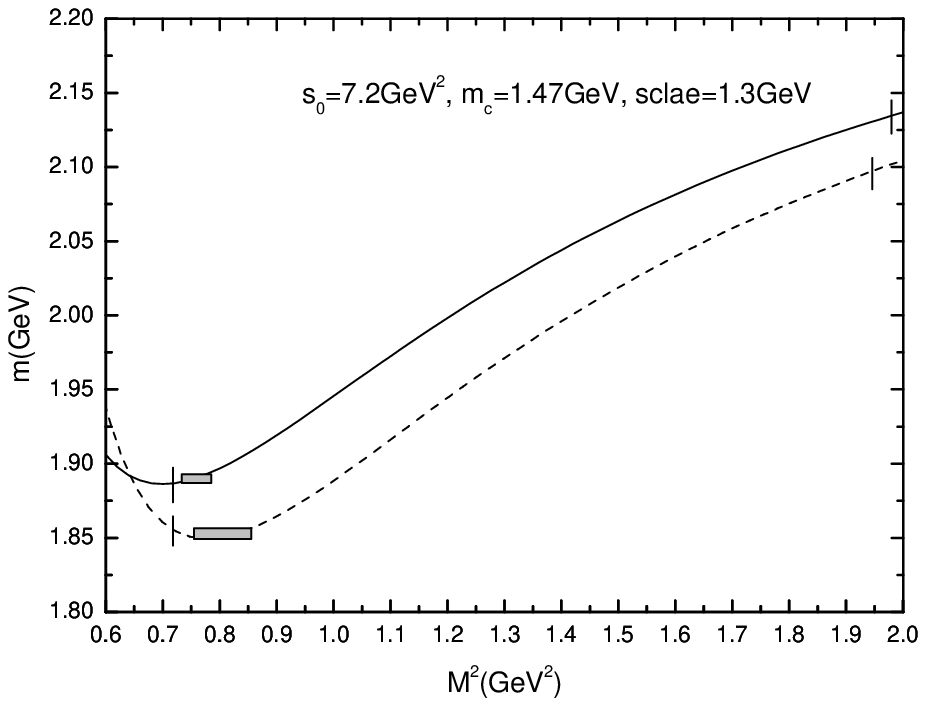}
\includegraphics[scale=0.75]{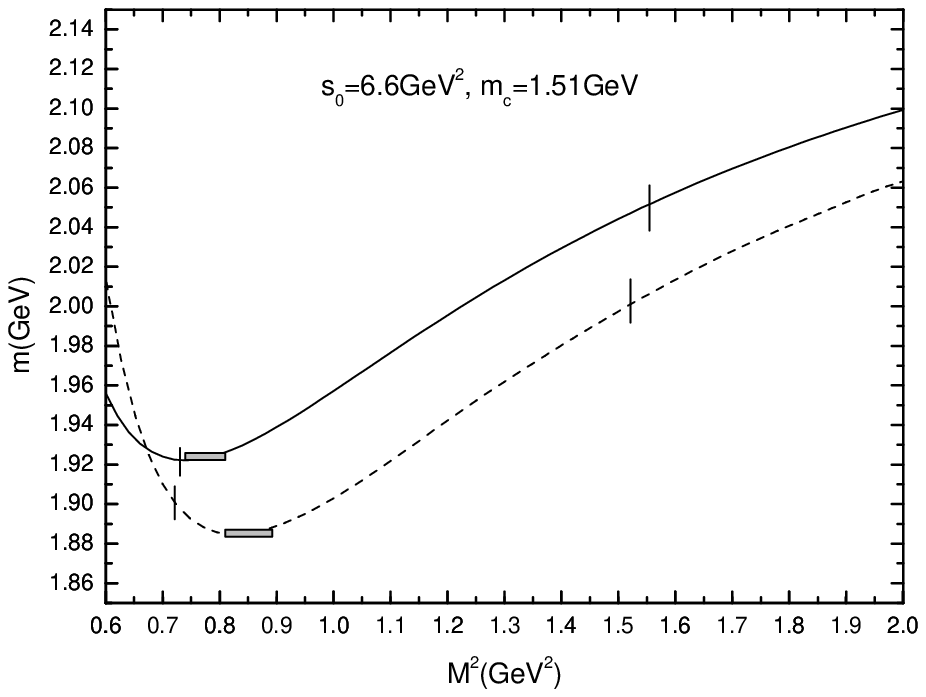}
\caption{Mass of $D_{d}$(dashed line) and $D_{s}$ meson (solid line)
of $0^{-}$ channel from pseudoscalar sum rule of Eq.(\ref{smr}) as
function of Borel momentum $M^{2}$ at scale $\mu=1.3\rm{GeV}$.}\label{psup}
\end{center}
\end{figure}

\begin{figure}
\begin{center}
\includegraphics[scale=0.75]{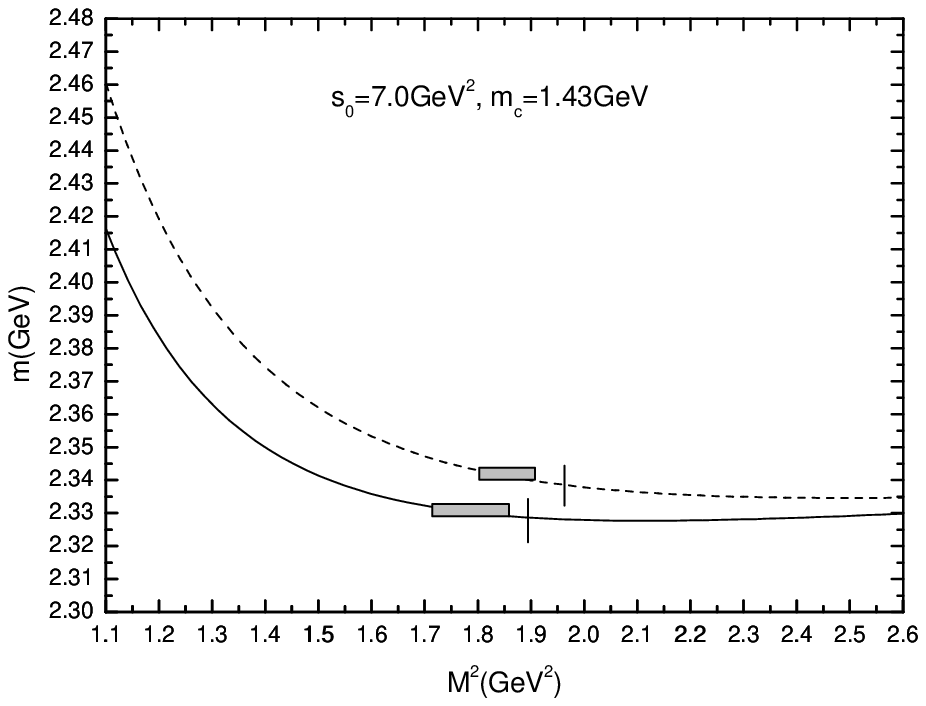}
\includegraphics[scale=0.75]{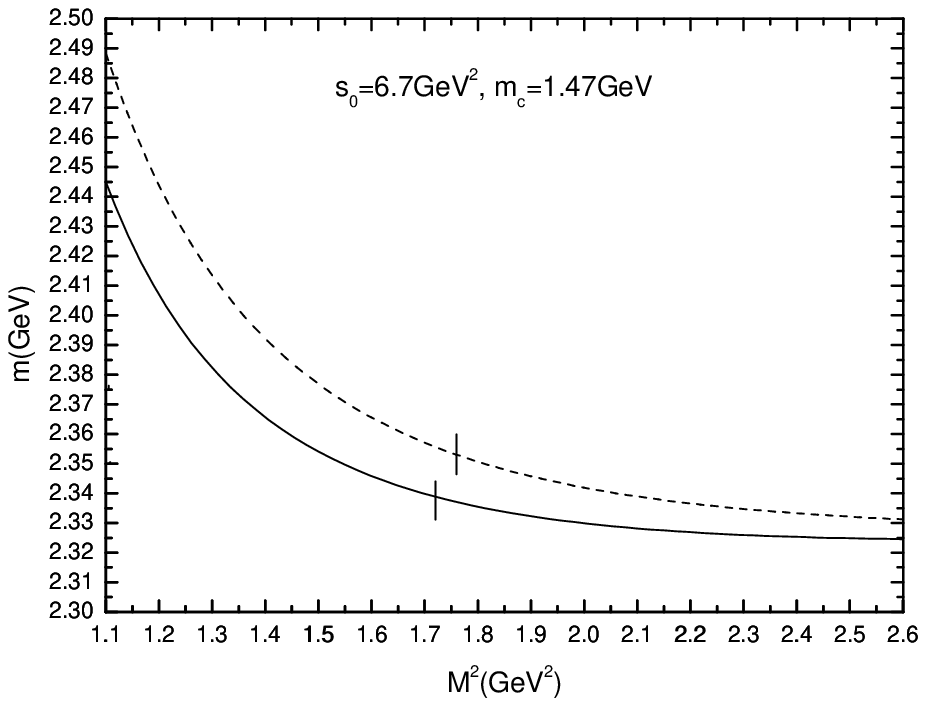}
\includegraphics[scale=0.75]{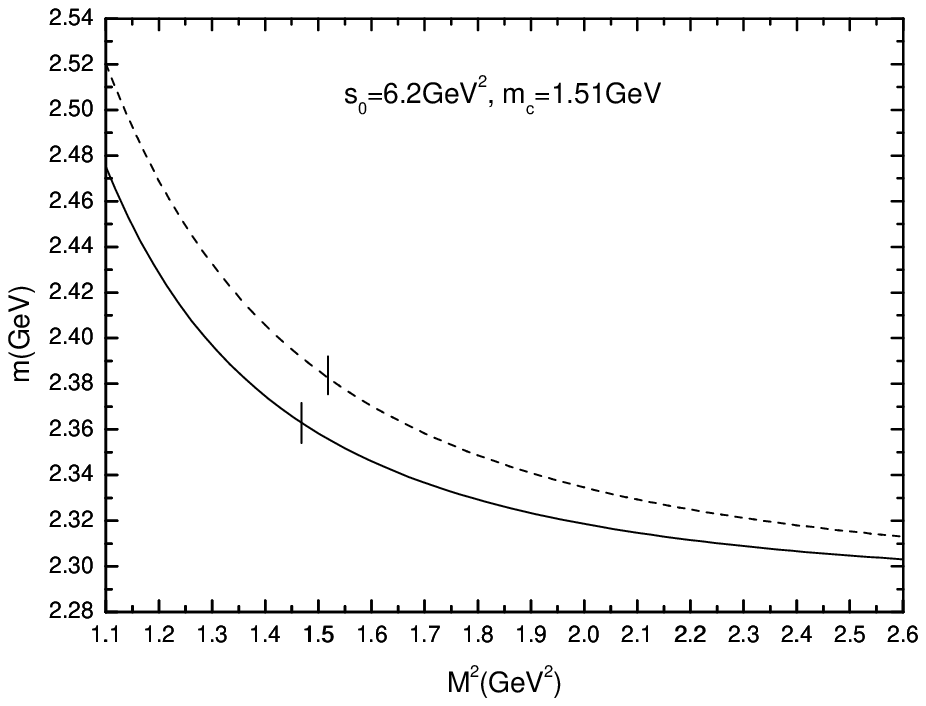}
\caption{Mass of $D_{d}$(dashed line) and $D_{s}$ meson (solid line)
from scalar sum rule of Eq.(\ref{smr}) as function of Borel momentum
$M^{2}$ at $\mu=1.3\rm{GeV}$. It is obvious with the charm mass
increasing, it is difficult to develop an reasonable extremum value
in the allowed Borel window. }\label{scalarup}
\end{center}
\end{figure}

\begin{table}[t]
\caption{Mass of pseudoscalar $D_{d}$ and $D_{s}$ read from shaded
area marked by short bars in figure.\,\ref{psup} for different
threshold $s_{0}$ and $m_{c}$ at scale $\mu=1.3\rm{GeV}$. }
\begin{ruledtabular}
\begin{tabular}{c c c c c}
$s_{0}({\rm{GeV^{2}}})$ & $m_{c}(\rm{GeV})$ & $D_{d}(\rm{MeV})$ & $D_{s}(\rm{MeV})$ & $M_{D_{s}}-M_{D_{d}}(\rm{MeV})$ \\
\hline
 $7.6$ & $1.43$ & $1807$ & $1845$ & $38$\\
$7.2$ & $1.47$ & $1851$ & $1886$  & $35$\\
$6.6$ & $1.51$ & $1885$ & $1922$  & $37$\\
\end{tabular}
\end{ruledtabular}\label{tabpsup}
\end{table}

It is instructive to study the scale dependence of our results since
physical quantities are scale-independent thus it will supply a
natural check on our results. Therefore the theoretical splitting
should be unchanged when  calculated with another scale. To this end
we evolve the related parameters according to Eq.(\ref{rge}) to a
higher scale $\mu=1.3\rm{GeV}$ which is still lower than charm mass.
The results for $0^{-}$ and $0^{+}$ are shown in figure.\,\ref{psup}
and figure.\,\ref{scalarup} respectively. We can see that when the
scale increases, the results of $0^{-}$ channel still keep a well
behavior as $\mu=1\rm{GeV}$. We can read from figure.\,\ref{psup}
the splitting $M_{D_{s}}-M_{D_{d}}\sim 35\rm{MeV}$ which agrees well
compared with $\mu=1\rm{GeV}$. But the situation is not so good in
the $0^{+}$ channel when we scale up. It is obvious from
figure.\,\ref{scalarup} that at $\mu=1.3\rm{GeV}$ the calculated
mass is monotonically decreasing within the window satisfying the
criterion 1 in $0^{+}$ channel. But fortunately the calculated mass
of $D_{d}$ is always larger than $D_{s}$ within the selected Borel
windows. For example, if we take the central values of in
figure.\,\ref{scalarup} we find the mass gap
$M_{D_{d}}-M_{D_{s}}\sim 15\rm{MeV}$ which is also consistent with
the splitting obtained at $\mu=1\rm{GeV}$. Therefore the splitting
in both channels are invariant which shows a correct scale
invariance. Combined the results at $\mu=1\rm{GeV}$ we conjecture
the reason why it is difficult to develop an extremum value at large
charm mass in $0^{+}$ channel maybe, as pointed out
in\cite{snarison}, is that a large charm mass will induce large
error. Our results imply that it is more appropriate to take a lower
pole mass for charm. In fact as the scale increases, we approach to
the asymptotic free side further thus the non-perturbative effects
will have reduced impact. We can see obviously from
figure.\ref{scalarup} that at the high energy side in the $0^{+}$
channel the mass gap decreases.

\begin{figure}
\begin{center}
\includegraphics[scale=1.0]{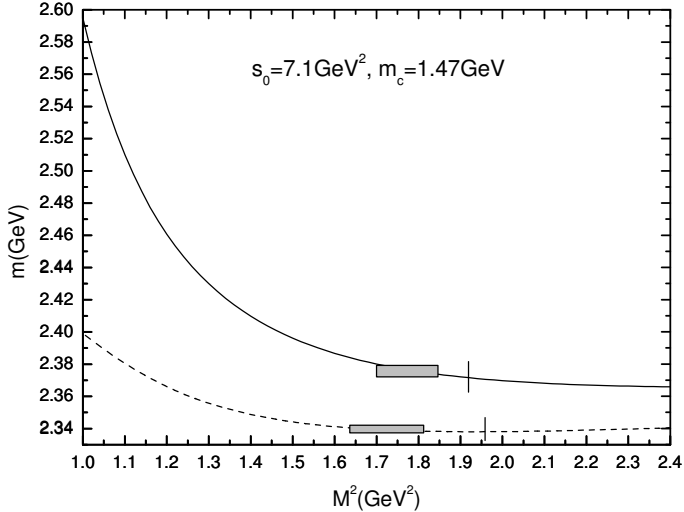}
\caption{Mass curves of $D_{d}$(dashed line) and $D_{s}$(solid line)
vs. Borel momentum $M^{2}$ in $0^{+}$ channel at $\mu=1\rm{GeV}$,
where we set $\langle\bar{d}d\rangle=\langle\bar{s}s\rangle$. It is
clear the splitting of the two states reverse.}\label{flip}
\end{center}
\end{figure}

The different importance of mass effects in realizing the splitting
in $0^{-}$ and $0^{+}$ channel of $D$ meson is not surprising. One
should notice the ``\,force\,'' induced by the QCD vacuum or
equivalently, the non-perturbative effects is parity-dependent which
is well indicated by the contribution of the dominant condensates
$m_{c}\langle\bar{q}q\rangle$ and $m_{c}\langle\bar{q}\sigma
Gq\rangle$ as well as the parts in
$\langle\alpha_{s}G^{2}/\pi\rangle$ introduced by their mixing. In
$0^{-}$ channel  the entire effect of these two terms give positive
contributions to the correlation function thus an attractive force
is  induced by the QCD vacuum. While in $0^{+}$ channel these two
terms supply negative contributions to the correlation function,
then a repulsive force is induced by the QCD vacuum. Since
$-\langle\bar dd\rangle>-\langle\bar ss\rangle$ we could expect
$M_{D_s}>M_{D_d}$ in $0^{-}$ and $M_{D_s}<M_{D_d}$ in $0^{+}$
channel. Furthermore the ``\,force\,'' is scale dependent which
implies that the larger the scale is, the farther we leave from the
confinement sector, therefore the importance of non-perturbative
effects will be discounted compared with a lower scale. In the
$0^{+}$ channel the effect of the quark condensate $\langle\bar
qq\rangle$ overpowers other mass effects, so we find
$M_{D_s}<M_{D_d}$. On the contrary if we set
$\langle\bar{d}d\rangle=\langle\bar{s}s\rangle$ then the mass
difference in $0^{+}$ channel is produced by $m_{q}$-dependent terms
only; thus it is expected there will be mass-flipping. The mass
curves are shown in figure.\,\ref{flip}. It is obvious that the mass
gap of the two scalars are very sensitive to the ratio
$\kappa=\langle\bar{s}s\rangle/\langle\bar{u}u\rangle$. However, in
$0^{-}$ channel the sign of the splitting remains unchanged, the
mass gap of the two states is not sensitive to this ratio.

In fact we can categorize corrections into two parts: one is
parity-dependent and mainly proportional to light quark masses,
$\langle\bar{q}q\rangle$ and $\langle\bar{q}\sigma Gq\rangle$,
another is parity-independent such as $m_q\langle\bar{q}q\rangle$,
$m_q\langle\bar{q}\sigma Gq\rangle$ and $\langle\bar{q}q\rangle^2$.
It seems the former overpower  the latter because the latter are
doubly suppressed by the $m_q$ and the $\langle\bar{q}q\rangle$,
however, their magnitudes for SU(3)-breaking are comparable. The
former change their signs when we alter from $0^{+}$ to $0^{-}$
channel and vice versa while the latter  do not. We can learn from
pseudoscalar sum rule in Eq.(\ref{smr}) that in the $0^{-}$ channel
these two parts provide a consistent response to the splitting since
there is flipping in their signs. For instance the mass gap from
quark condensates is:
\begin{eqnarray}
m_{c}(\langle\bar{s}s\rangle-\langle\bar{d}d\rangle)>0,\nonumber\\
m_{d}\langle\bar{d}d\rangle-m_{s}\langle\bar{s}s\rangle>0,\nonumber
\end{eqnarray}
and from the other condensates all contribute positive differences,
all these positive difference are inclined to broaden the mass gap
between $D_{s}$ and $D_{d}$ in $0^{-}$ channel. However, in the $0^{+}$
channel all the non-perturbative corrections keep the same positive
parity, and consequently a competitive role to the splitting of $D_{d}$
and $D_{s}$ developed:
\begin{eqnarray}
m_{c}(\langle\bar{d}d\rangle-\langle\bar{s}s\rangle)<0,\nonumber\\
m_{d}\langle\bar{d}d\rangle-m_{s}\langle\bar{s}s\rangle>0,\nonumber
\end{eqnarray}
so does the $\langle\bar{q}\sigma Gq\rangle$. This means there are
some compensations to the splitting from the parity-independent
terms which prefer to weaken the splitting between $D_{d}$ and
$D_{s}$ in $0^{+}$ channel. Therefore the mass gap is broader in the
$0^{-}$ channel than in $0^{+}$ of $D$ meson. It is worthy to
mention that this phenomenon is partly noticed in\cite{snarison}. In
fact these results can be generalized to the $0^{-}$ and $0^{+}$
channels of pure-light mesons, but the mass difference induced by
various condensates is greatly suppressed by small light quark mass
thus it is expected the splitting is tiny by this mechanism, so to
realize a realistic splitting in QCD sum rules based on a naive
quark model should take into account instanton effects\cite{shuryak,
zhang}.

\begin{figure}
\begin{center}
\includegraphics[scale=0.75]{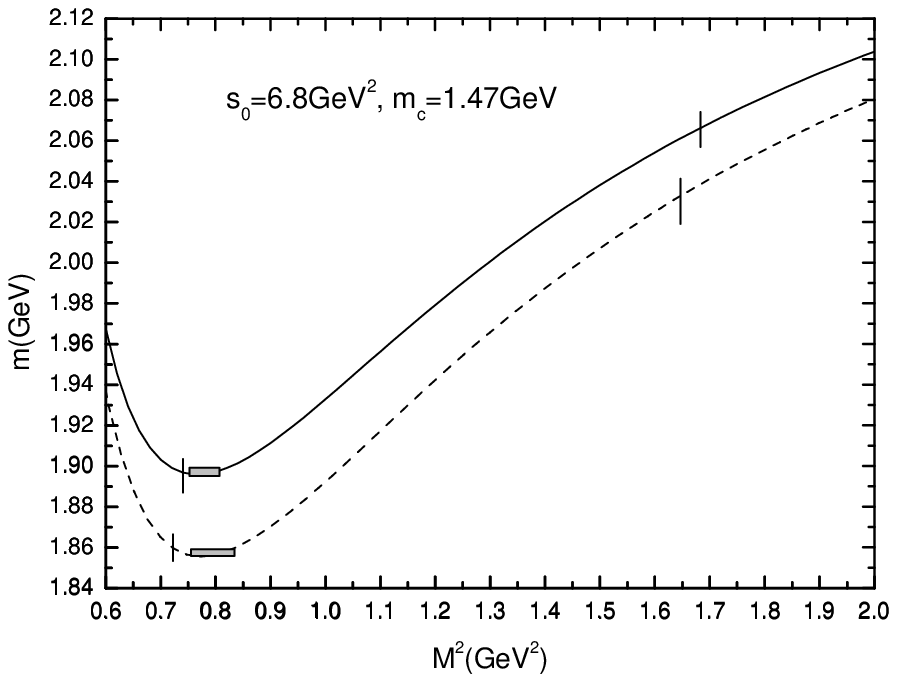}
\includegraphics[scale=0.75]{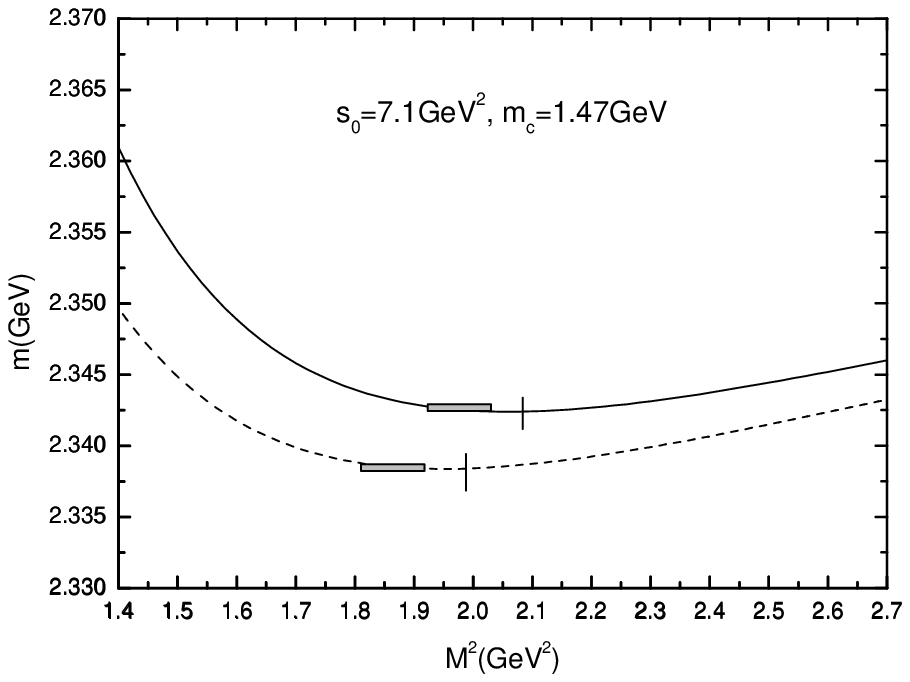}
\caption{Mass curves of $0^{-}$(left) and $0^{+}$(right) channels
vs. Borel momentum $M^{2}$ at $\mu=1\rm{GeV}$, where we have turned
off the $m_{q}$-dependent corrections in $C_{G^{2}}$ in sum rules.
The dashed and solid line in both graphs denote $D_{d}$ and $D_{s}$
respectively. }\label{nomq}
\end{center}
\end{figure}

As operator mixing changes the coefficient of two-gluonic
condensates significantly, we can see the $m_{q}$-dependent parts in
$C_{G^{2}}$ are parity-dependent and have a complicated form but
also an obvious scale-dependence. It is instructive to investigate
the impact on the calculated mass of both channels. For this purpose
we turn off $m_{q}$-dependent parts in $C_{G^{2}}$ in sum rules. The
mass curves are shown in figure.\ref{nomq}. It is obvious that in
this case in both channels $M_{D_{s}}$ is larger than $M_{D_{d}}$.
We find in $0^{-}$ channel the mass gap is broader than there is
$m_{q}$-dependent corrections in gluonic condensate $\sim
\rm{10MeV}$ so our results show this correction is negative in the
mass gap for the $0^{-}$ channel. In $0^{+}$ channel the situation
is just opposite to the case when $m_{q}$-dependent corrections turn
on, thus the results show this contribution is positive to realize a
reasonable splitting in $0^{+}$ channel.

A natural idea is to generalize these arguments to the $1^{-}$ and
$1^{+}$ channel of $D$. Unfortunately it does not work which can be
well understood from heavy quark effective theory\cite{hqet}. When
there is no orbital excitation of light content, the c-quark with
spin $s_{c}=1/2$ and the light degrees of freedom with spin
$s_{l}=1/2$ forming a multiplet of hadrons with spin:
\begin{equation}
j=\frac{1}{2}\otimes\frac{1}{2}=0\oplus1,\nonumber
\end{equation}
thus it is clear there is a unique $J^{P}=1^{-}$ multiplet. While if
there is orbital excitation of light degrees of freedom, the spin of
light content will be:
\begin{equation}
s_{l}=l\pm\frac{1}{2}=\frac{1}{2}\quad{\rm{or}}\quad\frac{3}{2},\nonumber
\end{equation}
combining with the spin of c-quark $s_{c}=1/2$:
\begin{equation}
j=\frac{1}{2}\otimes\frac{1}{2}=0\oplus1,\nonumber
\end{equation}
\begin{equation}
j=\frac{1}{2}\otimes\frac{3}{2}=1\oplus2.\nonumber
\end{equation}
so there will be two $1^{+}$ multiplets, the $D_{1}^{\ast}$ and the
$D_{1}$ states experimentally. These two states are very close, thus
the single resonance approximation in QCD sum rules is not viable.
 If we let the charm mass go to infinity these two $1^{+}$ states can
be separated in the formalism of the heavy quark effective
theory\cite{dai}. However, the $1/m_c$ corrections which are the
same order of SU(3)-breaking effects make these two states mix
again. So we still cannot get a simple correspondence between
$1^{-}$ and  $1^{+}$. We can resort to the experimental data
directly\cite{amsler}. The mass splitting between $D_s$ and $D_d$ in
$1^{-}$ is about 100MeV, while it is  only about 40MeV in the lower
$1^{+}$(which corresponds to $1^{-}$ in the heavy quark mass limit).
A similar effect still appears.

Finally let us briefly mention the $B$ case although there is not
enough experimental evidence. Since the b-quark mass is so large
 the SU(3)-breaking effects are smeared in the formalism
Eq.(\ref{smr}). We can hope the similar effects will be recovered in
the formalism of the heavy quark effective theory. Certainly,
SU(3)-breaking effects also appear in $1/m_h$ corrections which
might not be small and could cause some differences between $D$
mesons and $B$ mesons.

\section{conclusions}

In this work, based on the pseudoscalar and scalar sum rule from
$c\bar{q}$ structure we investigate the SU(3)-breaking effects
enhanced by the large charm mass on the splitting of the
pseudoscalar and scalar $D$ multiplet. Since the quark condensates
$\langle\bar{q}q\rangle$ and $\langle\bar{q}\sigma Gq\rangle$ are
greatly enhanced by the heavy quark mass, they play more important
roles in the SU(3)-breaking in the heavy-light mesons. The sign of
the $\langle\bar{q}q\rangle$ and $\langle\bar{q}\sigma Gq\rangle$
contributions is different in $0^{-}$ and $0^{+}$ channels thus
resulting a ``\,parity- dependent\,'' force which can explain the
relatively-lower mass of $D_{s}(2317)$. Furthermore, these
parity-dependent corrections broaden the mass gap in the $0^{-}$
channel while weakening it in $0^{+}$ channel. The results show that
the $\mathcal {O}(m_{q})$ corrections in two-gluonic condensates
introduced by operator mixing are noticeable in $0^{+}$ channel.
Such a ``\,parity-dependent\,'' force should exist generally in the
heavy-light mesons, but its magnitude should be dependent on the
specific system. This force is also energy-dependent, so its effect
should be weaker in high excited states. We also analyze the cases
$1^{-}$ and $1^{+}$. The experimental data hints such a force also
exists but is weaker than that in $0^{-}$ and $0^{+}$.

\section{Acknowledgements}
This work is supported by NNSFC under Projects No. 10775117,
10847148. The author H.Y. Jin  thanks Hai-Yang Cheng for very useful
discussion.

\begin{appendix}

\section{derivation of Eq.(\ref{finalGGG})}
It is a little effort to work out $C_{G^{3}}$ in scalar current
expansion from a vector current expansion. To this end, let us
consider vector current two-point function:
\begin{eqnarray}
\Pi_{\mu\nu}(q^{2})&=&i\int d^{4}xe^{iqx}\langle
0|T\{\bar{q}(x)\gamma_{\mu}c(x),
\bar{c}(0)\gamma_{\nu}q(0)\}|0\rangle\nonumber\,\\
&=&(-g_{\mu\nu}q^{2}+q_{\mu}q_{\nu})\Pi^{\rm{V}}(q^{2})+q_{\mu}q_{\nu}\Pi^{\rm{S}}(q^{2}),\label{vectorexpan}
\end{eqnarray}
To single out the scalar part we contract Eq.(\ref{vectorexpan})
with $q^{\mu}q^{\nu}$:
\begin{equation}
q^{\mu}q{^\nu}\Pi_{\mu\nu}(q^{2})=q^{4}\Pi^{\rm{S}}(q^{2}),\label{scalarpart}
\end{equation}
Then it is convenient to consider the following two-point function
based on the four-divergence of vector current:
\begin{eqnarray}
\Pi^{\rm{S'}}(q^{2})&=&i\int d^{4}xe^{iqx}\langle
0|T\{\partial^{\mu}[\bar{q}(x)\gamma_{\mu}c(x)]\partial^{\nu}[\bar{c}(0)\gamma_{\nu}q(0)]\}|0\rangle\nonumber\,\\
&=&i\int d^{4}xe^{iqx}\langle
0|\Big\{\bar{q}(x)\overleftarrow{\slashed{D}}c(x)
+\bar{q}(x)\overrightarrow{\slashed{D}}c(x)\Big\}\times\Big\{\bar{c}(0)\overleftarrow{\slashed{D}}q(0)
+\bar{c}(0)\overrightarrow{\slashed{D}}q(0)\Big\}|0\rangle\nonumber\,\\
&=&i(m_{c}-m_{q})^{2}\int d^{4}xe^{iqx}\langle
0|\bar{q}(x)c(x)\bar{c}(0)q(0)|0\rangle,\label{divexpan}
\end{eqnarray}
where the Dirac equations:
\begin{equation}
\overrightarrow{\slashed{\partial}}\psi(x)=-im\psi(x),\nonumber
\end{equation}
and
\begin{equation}
\bar{\psi}(x)\overleftarrow{\slashed{\partial}}=im\bar{\psi}(x).\nonumber
\end{equation}
have been used. In this way we can associate OPE of scalar current
expansion with the vector one:
\begin{equation}
i\int d^{4}xe^{iqx}\langle
0|\bar{q}(x)c(x)\bar{c}(0)q(0)|0\rangle=\frac{\Pi^{\rm{S'}}(q^{2})}{(m_{c}-m_{q})^{2}},\label{relation}
\end{equation}
Combined with Eq.(\ref{scalarpart}) it is straightforward to obtain:
\begin{equation}
i\int d^{4}xe^{iqx}\langle
0|\bar{q}(x)c(x)\bar{c}(0)q(0)|0\rangle=\frac{q^{4}\Pi^{\rm{S}}(q^{2})}{(m_{c}-m_{q})^{2}},\label{combination}
\end{equation}
And $C_{G^{3}}$ has been worked out\cite{scgeneralis} in heavy-light
vector current expansion, here we write it in a standard form:
\begin{eqnarray}
C_{\mu\nu}^{G^{3}}(q^{2})&=&(-g_{\mu\nu}q^{2}+q_{\mu}q_{\nu})A+g_{\mu\nu}(m_{c}-m_{q})^{2}\nonumber\,\\
&=&(-g_{\mu\nu}q^{2}+q_{\mu}q_{\nu})\Big(A-\frac{(m_{c}-m_{q})^{2}}{q^{2}}B\Big)
+q_{\mu}q_{\nu}\Big(-\frac{(m_{c}-m_{q})^{2}}{q^{2}}\Big)B.\label{vectorGGG}
\end{eqnarray}
where
\begin{equation}
A=\frac{W^{2}}{4320\pi^{2}m_{c}^{6}}(-30W^{2}-8W-3),\nonumber
\end{equation}
\begin{equation}
B=\frac{W}{2880\pi^{2}m_{c}^{6}}(-10W^{3}+4W^{2}+3W+2),\nonumber
\end{equation}
\begin{equation}
W=\frac{m_{c}^{2}}{m_{c}^{2}-q^{2}}.\nonumber
\end{equation}
Combined with Eq.(\ref{combination}) we can obtain $C_{G^{3}}$ in a
scalar heavy-light expansion:
\begin{equation}
C_{G^{3}}=-\frac{q^{2}}{720m_{c}^{6}}W(-10W^{3}+4W+3W+2).\label{final}
\end{equation}
This is Eq.(\ref{finalGGG}) where we have suppressed the factor
$\frac{\alpha_{s}}{\pi}$.

\end{appendix}

\end{document}